\documentclass[11pt,twoside]{article}
\usepackage{asp2010}

\usepackage{amsmath}

\resetcounters

\begin{document}

\title{Direction Dependent Effects In Wide-Field Wideband Full Stokes Radio Imaging}
\author{Preshanth Jagannathan$^{1,2}$, Sanjay Bhatanagar$^1$, Urvashi Rau$^1$ and Russ Taylor$^{2,3}$
\affil{$^1$National Radio Astronomy Observatory, 1003, Lopezville Road, Socorro, 87801, New Mexico, U.S.A}
\affil{$^2$Department of Astronomy, University of Cape Town, Private Bag X3, Rondebosch 7701, Republic of South Africa}
\affil{$^3$Department of Physics, University of Western Cape, Modderdam Road, Private Bag X17,
Bellville, 7530, Republic of South Africa}}

\begin{abstract}
Synthesis imaging in radio astronomy is affected by instrumental and atmospheric effects which introduce direction-dependent (DD) gains.The antenna power pattern varies both as a function of time and frequency. The broad band time varying nature of the antenna power pattern when not corrected leads to gross errors in full Stokes imaging and flux estimation. In this poster we explore the errors that arise in image deconvolution while not accounting for the time and frequency dependence of the antenna power pattern. Simulations were conducted with the wide-band full Stokes power pattern of the Karl G. Jansky Very Large Array (VLA) antennas to demonstrate the level of errors arising from direction-dependent gains and their non-neglegible impact on upcoming sky surveys such as the VLASS. DD corrections through hybrid projection algorithms are computationally expensive to perform. A highly parallel implementation through high performance computing architectures is the only feasible way of applying these corrections to the large data sizes of these upcoming surveys.\end{abstract}

\section{Measurement Equation}
To achieve high dynamic range imaging in interferometry corrections for the variations of antenna gain over the primary beam of the antennas will be required.  These so-called direction dependent (DD) gains can be expressed using the formulation for direction independent (DI) gains of  \cite{1996A&AS117137H}
\begin{equation*}
\centering
\begin{pmatrix} S_{pp} \\ S_{pq} \\ S_{qp} \\ S_{qq} \end{pmatrix}_{Obs} = \begin{pmatrix}J_{i}^{p}J_{j}^{p*} & -J_{i}^{p}J_{j}^{pq*} & -J_{i}^{pq}J_{j}^{p*} & J_{i}^{pq}J_{j}^{pq*}\\
J_{i}^{p}J_{j}^{qp*} & J_{i}^{p}J_{j}^{q*} & -J_{i}^{pq}J_{j}^{qp*} & -J_{i}^{pq}J_{j}^{q*}\\
J_{i}^{qp}J_{j}^{p*} & -J_{i}^{qp}J_{j}^{pq*} & J_{i}^{q}J_{j}^{p*} & -J_{i}^{q}J_{j}^{pq*}\\
J_{i}^{qp}J_{j}^{qp*} & J_{i}^{qp}J_{j}^{q*} & J_{i}^{q}J_{j}^{qp*} & J_{i}^{q}J_{j}^{q*}
\end{pmatrix} \begin{pmatrix} S_{pp} \\ S_{pq} \\ S_{qp} \\ S_{qq} \end{pmatrix}_{Sky} \
\end{equation*}

\begin{equation}
= \begin{pmatrix}M_{00} & M_{01} & M_{02} & M_{03}\\
M_{04} & M_{05} & M_{06} & M_{07}\\
M_{08} & M_{09} & M_{10} & M_{11}\\
M_{12} & M_{13} & M_{14} & M_{15}
\end{pmatrix}\begin{pmatrix} S_{pp} \\ S_{pq} \\ S_{qp} \\ S_{qq} \end{pmatrix}_{Sky}
\label{eqn:Mueller}
\end{equation}

\begin{figure}
\centering
 \includegraphics[width=0.9\linewidth]{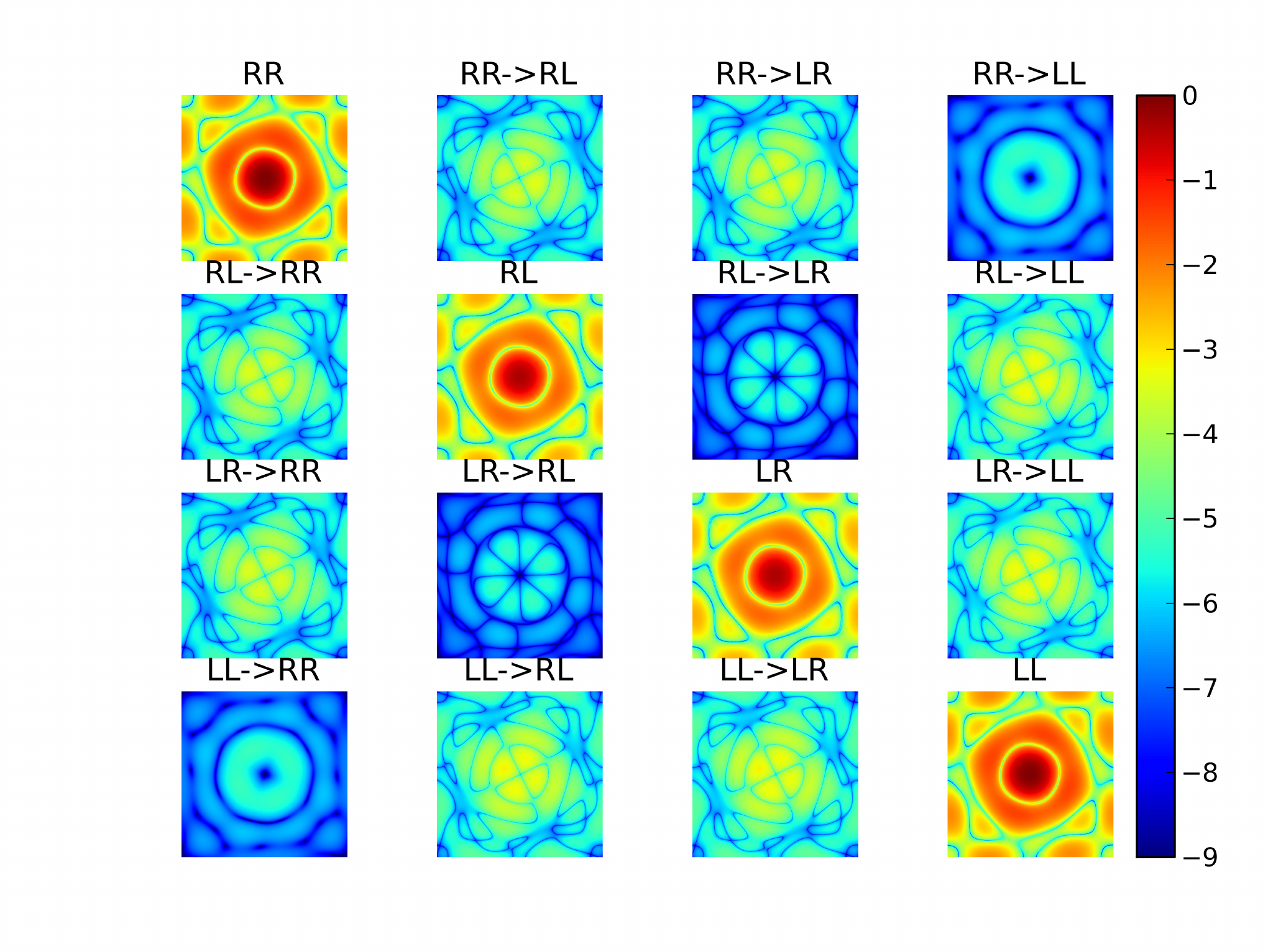}
 \caption{The plot above is the diagonal normalised amplitude of the complex
Mueller matrix in circular feed basis for the L-Band of the 
VLA at 1.064GHz. The normalised logarithmic colour scale has the peak flux of the $R$ and $L$ beams in the beam Jones matrix set to unity.}
\label{fig:Mueller}
\end{figure}

\noindent The symbols used in the equations above follow those in
\cite{BhatnagarDD2008A&A487419B}.  The elements of the Mueller matrix
($M_{ij}$) in the equations above are images, shown in
Fig.~\ref{fig:Mueller} for the VLA antennas, representing the
instrumental response to various polarization products; the in-beam DD leakage
terms.  The dominant terms are along the diagonal, while the weakest
terms are along the anti-diagonal (representing DD leakage in
orthogonal polarization products).  The magnitude of the other
off-diagonal elements$(10^{-3} - 10^{-9})$ show that errors due to ignoring these terms during
imaging will be significant beyond an imaging dynamic range of $\sim
10^{3-4}$ in Stokes-I.  However, since the Stokes-Q and -U signal is
typically at the level of a few percentage of Stokes-I, these errors
can degrade the Stokes-Q and -U image fidelity (and possibly also
dynamic range) by as much as 100\%.  Full-Mueller imaging is therefore
critical for noise-limited wide-field full-polarization
imaging with most modern telescopes.  Extending the Wide-band
A-Projection algorithm \citep{BhatnagarWBAWP2013ApJ77091B} for
full-Mueller imaging to account for in-beam time-, frequency- and
polarization dependent DD gains is therefore required to fully exploit
the scientific capabilities of modern radio telescopes.

\section{Simulation and Results}

\begin{figure}
\centering
 \includegraphics[width=0.9\linewidth]{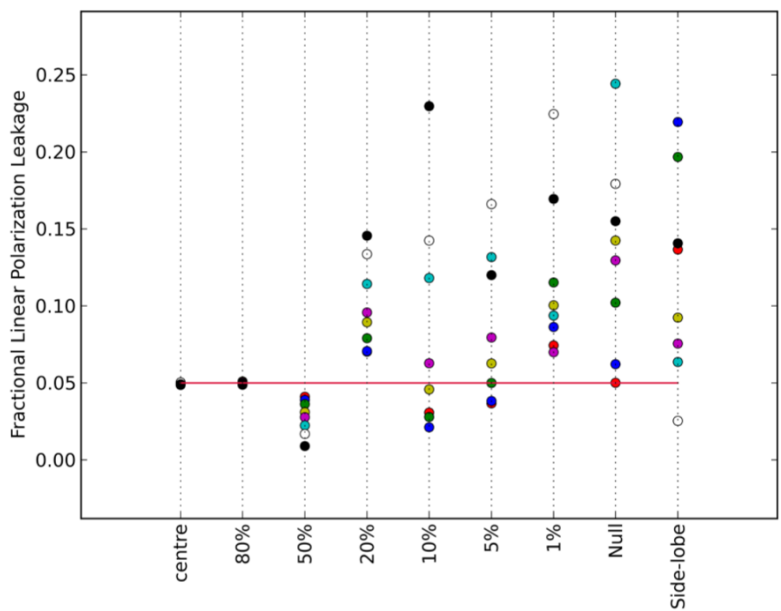}
 \caption{Plotted above is the fractional polarization ($p=\sqrt{Q^2 + U^2}/I$) of the nine points in the simulation. The known fractional polarization of each source is $p=0.05$ (red line to indicate the input fractional polarization). The colours of the points indicate the central frequency of the sub-band used in the simulation.R - $1.064$ GHz, B - $1.192$~GHz, G - $1.32$ GHz, Y - $1.448$ GHz, M - $1.576$ GHz, C - $1.704$ GHz, W - $1.832$ GHz,Bl - $1.96$ GHz}
 \label{fig:fracpol}
\end{figure}

To quantify the effect of the off-diagonal DD terms for the VLA we have carried out imaging simulations. A true sky image cube containing nine points of equal flux density [I=1.0,Q=0.025,U=0.0433,V=0] Jy in each of its channels, is convolved with the Mueller matrix from which visibilities are predicted per baseline, per parallactic angle. The combined data is then imaged by means of the CLEAN task in CASA using Clark CLEAN and standard MFS (\cite{2011MSMFS}) to produce an image cube containing $8$ channels of $128$~MHz, spanning $1$~GHz in bandwidth across the L-Band of the VLA from $1-2GHz$. The simulations ran from $-4h$ to $+4h$ in hour angle which ensures equal hour angle coverage across the meridian crossing. This is essential as an unequal hour angle coverage across the meridian would introduce a change in the source polarization position angle (PPA) . While the change in the PPA can be accounted for by ensuring equal hour handle coverage across the meridian, it however still does not account for the flux leakage given by the Mueller matrix. Plotted as a function of position in the beam in Figure~\ref{fig:fracpol} is the fractional polarization of the sources in the image. The plot shows in the induced polarization as a function of frequency (colour of the points) and as a function of beam position. As we move beyond the $80\%$ power in the beam the induced polarization reduces the polarization to $~50\%$ of the original at HPBW. Beyond half power the induced polarization continues to rise to an excess of well over $~100\%$ by the $20\%$ power in the beam. Therefore in the absence of full Mueller imaging the fidelity (flux accuracy) of the image is in question beyond the $~80\%$ power at the reference frequency of the cube image.

\section{Wideband Full Stokes AW Projection and Future Work}
Our work shows that wife field imaging in full stokes will require corrections for DD effects.
Such effects can be corrected for using the wide-band A-Projection algorithm \citep{BhatnagarWBAWP2013ApJ77091B}
to account for the time and frequency dependent effects of the antenna primary beam. \cite{2013LOFAR} has shown the effectiveness of
the A-Projection algorithm in Stokes-I imaging for aperture-array
telescope where the off-diagonal Mueller terms are as much as
$\sim10\%$ of the diagonal. The computational load of a full-Mueller
imaging depends on (i) the support size of the convolution function,
and (ii) the number of Mueller terms included.  A full Mueller matrix
reduction would naively imply $16\times$ the number of convolution
function calculations during the gridding and de-gridding
process. However for antenna-array telescope, short-cuts might be
possible.  Even then, this can be prohibitively expensive for large
datasets.  The computing however is good fit for parallel computing
and a parallel computing framework for using A-Projection in CASA is
currently under scientific testing. In the near future we hope to
employ and test a Wide-band full-Stokes A-Projection algorithm
utilising the DD Mueller matrix framework in CASA for deep radio polarization,
of wide-field, mosaic imaging(e.g. \cite{2014arXiv1405.0117T})
 
 \section{Conclusions}
 \begin{itemize}
  \item High dynamic range imaging in stokes $>10000$ requires wideband full stokes Mueller matrix corrected imaging by means of Wideband A-Projection.
  \item Polarization even for modest dynamic ranges of $>1000$ requires full Mueller matrix corrections as the leakage beyond the $80\%$ power is comparable to the source polarized signal.
  \item High dynamic range and high fidelity are two different criteria for image quality, unless DD effects are handled properly high dynamic range does not translate to high fidelity 
\end{itemize}

\acknowledgements 
Preshanth Jagannathan is a Reber Pre-Doctoral Fellow at the National Radio Astronomy Observatory and is supported by Associated Universities, Inc./National Radio Astronomy Observatory and the National Science Foundation.
\bibliography{P04-9}

\end{document}